\documentclass[reprint,unsortedaddress,superscriptaddress,footinbib,amsmath,amssymb,aps,pra,showkeys,showpacs]{revtex4-2}
\usepackage{graphicx}
\usepackage{xcolor}
\usepackage{amsmath}
\usepackage{amssymb}
\usepackage{braket}
\setcitestyle{numbers,super}

\begin{document}

\title{Quantum Entanglement Response to Pulsed Gate Modulation}

\author{E. M. Fernandes}
\email{ellen.fernandes@ufu.br}
\affiliation{Instituto de F\'{i}sica, Universidade Federal de Uberl\^andia,
38400-902, MG, Brazil}
\author{L. Sanz}
\email{lsanz@ufu.br}
\affiliation{Instituto de F\'{i}sica, Universidade Federal  de Uberl\^andia,
38400-902, MG, Brazil}
\author{F. M. Souza}
\email{fmsouza@ufu.br}
\affiliation{Instituto de F\'{i}sica, Universidade Federal de Uberl\^andia,
38400-902, MG, Brazil}

\date{\today}

\begin{abstract}
We examine the impact of time-dependent gate voltages on entanglement generation in two capacitively coupled charge qubits, with single-electron injection triggered on demand. The gate voltage modulates the tunnel coupling between the qubits and electronic reservoirs, initiating charge transport into the system. The formation of entangled states arises from the competition between inter-qubit Coulomb interactions and electron hopping processes. Particular attention is paid to the temporal structure of the gate pulse, which plays a pivotal role in shaping the entanglement dynamics. By exploring a variety of pulse profiles, we uncover regimes of enhanced entanglement and identify optimal driving conditions. Additionally, we investigate how environmental dephasing deteriorates entanglement formation. Within the framework of the density matrix formalism, we calculate fidelity, linear entropy, and negativity to identify robust operational windows. These results provide insights into controlling quantum correlations in mesoscopic systems and underscore the importance of error mitigation strategies in realizing high-performance electronic quantum devices.
\end{abstract}

\maketitle

\section{Introduction}
\label{sec:intro}

Quantum computation technologies have received growing attention in the last few years due to their potential to revolutionize information processing~\cite{nielsenchuang2007}. 
A few examples of algorithms include factoring~\cite{shor1994, shor1999}, database search~\cite{grover1996}, data fitting optimization~\cite{wiebe2012}, and quantum machine learning~\cite{biamonte2017}, all of which are fields where quantum computers could have a significant impact. However, to achieve those practical implementations some challenges should be overcome, such as suppression of noise sources and improvement of error correction circuits~\cite{leon2021}.

In the pursuit of efficient quantum computing hardware, a diverse array of technologies has emerged. For instance, it has been demonstrated the viability of a programmable superconducting circuits to create states of 53 qubits, where each qubit is coupled to four nearest neighbors in a rectangular array~\cite{arute2019}. Additionally, superconducting quantum processors with 127 qubits that runs quantum circuits of two-qubit gates have already shown advantages in the present technological status that lack fault-tolerant quantum circuits~\cite{ykim2023}. Also, superconducting qubits have been fabricated using industry-standard techniques on silicon wafers (CMOS manufacturing)~\cite{vandamme2024}. in addition to protocols to mitigate qubit decoherence errors~\cite{Rower2024}.

Alternatively, quantum computing based on semiconductors can also be of great significance as it relies on advanced semiconductor manufacturing~\cite{zwerver2022, chatterjee2021}.
Semiconductor quantum dot qubits, in particular, represent a promising system for quantum computing~\cite{Zhang2019}, with the qubits being possibly defined in a few different ways, such as singlet-triplet qubit~\cite{levy2002}, exchange-only qubit~\cite{divincenzo2000}, charge qubit~\cite{hayashi2003,shinkai2007, shinkai2009a, peterson2010, dkim2015}, and spin 1/2 qubit~\cite{takeda2016, yoneda2018, zajac2018,BURKARD2023}, with the latter two being relatively easy to control via gate voltages. 

On one hand, spin-based quantum computing with quantum dots has garnered significant attention since the pioneering work of Loss and DiVincenzo~\cite{loss1998}, which is favored by relatively long decoherence times~\cite{chatterjee2021}, also both one- and two-qubit gates with fidelity exceeding 99\% have been demonstrated~\cite{watson2018, mills2022}. However, spin states in quantum dots can only been measured via average signal from an ensemble of electron spins~\cite{ciorga2001,fujisawa2002} or individually with a single-shot read-out in a scheme based on spin-to-charge conversion~\cite{elzerman2004,connors20}. On the other hand, charge qubits in semiconductor quantum dots can be straightly manipulated and readout through gate voltages and electron transport~\cite{macquarrie20}. Yet, charge qubits can suffer more influence by the environment's charge fluctuations and electric field variations, which results in relatively shorter decoherence times, ranging from hundreds of picosecond up to a few nanoseconds depending on the charge qubit system~\cite{shi2013, dkim2015, UDDIN2022}.

Quantum entanglement is of fundamental importance for quantum computing~\cite{nielsenchuang2007},as it is a necessary resource for performing operations in universal quantum computation~\cite{divincenzo1995}.
The entanglement of two charge qubits can be of great relevance for semiconductor nanoelectronics based quantum computers~\cite{shinkai2009a, fujisawa2011, oliveira2015}. Here we focus on the entanglement formation between a pair of qubits within two double quantum dots (DQD) structure~\cite{shinkai2009a}. 
As the electronic initialization process is a crucial step in the quantum dynamics, we pay particular attention on how the gate pulse, that controls charge injection, can affect the entanglement of the qubits. Our main goal is to optimize the initialization gate 
parameters to achieve the highest entangled state as possible, thus providing further insight for experimental implementations of entangled charge qubits.

The structure of this paper is as follows. In Sec.~\ref{sec:theory}, we describe the system of interest and present the theoretical formalism, including detailed modeling of the coupling between the qubit pair and the environment, as well as the gate pulse shapes. Sec.~\ref{sec:results} focuses on the essential conditions required to achieve the ``sweet spot'' for generating an entangled state of electrons in our physical system. Finally, in Sec.~\ref{sec:conc}, we conclude by summarizing our findings.

\section{Theoretical formalism}
\label{sec:theory}
\subsection{Isolated Qubits}

The system analyzed in this study is illustrated in Fig.~\ref{system}. It consists of four quantum dots (QD) ($1$ to $4$) organized into two pairs ($1-2$ and $3-4$), with each pair featuring two dots coupled by coherent tunneling $\tau$. These quantum dots are connected to electronic reservoirs: the left and right leads (represented in yellow and gray) inject electrons into specific dots within the arrangement. Incoherent tunneling rate $\Gamma(t)$ between the leads and the dots is time-dependent, enabling charge injection to be performed on demand in a time-controlled manner. Our model does not consider the spin degree of freedom of the electron, in agreement with the pseudo-spin model used by the pioneering works in Refs.~\cite{hayashi2003,shinkai2007,shinkai2009a}. Additionally, electrons within the quantum dots interact via Coulomb repulsion.

We model our system using the following Hamiltonian,
\begin{equation}\label{H0}
        H_0 = h_{0} + v + h_{c},
\end{equation}
where $h_{0} = \sum_{i=1}^{4} \varepsilon_i \hat{n}_i$, with $\varepsilon_{i}$ representing the energy level of the $i$-th quantum dot and $\hat{n}_i = d_i^\dag d_i$ being the number operator. Here, the operator $d_i^\dag$ ($d_i$) corresponds to the creation (annihilation) of an electron in the $i$-th state. The electronic coherent hopping between neighboring dots is described by
\begin{equation}
	v = \gamma ( d_1^\dag  d_2 + d_3^\dag  d_4+ h.c).
\end{equation} 
Note that QD1 and QD2 hybridize their orbitals, forming a molecular structure that operates as a charge qubit in the current model. Similarly, QD3 and QD4 constitute a second qubit. Charge transfer is restricted between the qubits, i.e., there is no charge flow between the pairs (QD1, QD2) and (QD3, QD4). However, the qubits can interact capacitively through Coulomb repulsion, which is described by
\begin{equation}
	h_c=J (n_1 n_3 + n_2 n_4) + J'(n_1  n_4 + n_2 n_3),
\end{equation}
 with $J$ and $J'$ being the direct and crossed Coulomb interaction strengths.

\subsection{Fermion-to-qubit Mapping}

As we are intended to explore quantum entanglement properties, it becomes more convenient to write the second quantized operators as a spin-tensor array. Using a fermion-to-qubit mapping\cite{souza2017}, we have
\begin{eqnarray}
d_1 &=& \sigma_{-}\otimes\mathbb{I}\otimes\mathbb{I}\otimes\mathbb{I} \nonumber\\
d_2 &=& \sigma_{z}\otimes\sigma_{-}\otimes\mathbb{I}\otimes\mathbb{I} \\
d_3 &=& \sigma_{z}\otimes\sigma_{z}\otimes\sigma_{-}\otimes\mathbb{I} \nonumber\\
d_4 &=& \sigma_{z}\otimes\sigma_{z}\otimes\sigma_{z}\otimes\sigma_{-} \nonumber
\end{eqnarray}
where $\sigma_-=(\sigma_x-i\sigma_y)/2$ and $\sigma_+= (\sigma_x+i\sigma_y)/2$, with $\sigma_ x$, $\sigma_ y$, $\sigma_ z$ being the Pauli matrices. 
The creation operators $d_i^\dagger$ are simply the conjugate transpose of $d_i$. This representation provides a clearer computational basis, expressed as $\{\ket{0000}$, $\ket{0001}$, $\ket{0010}$, $\ket{0011}$, $\ket{0100}$, $\ket{0101}$, $\ket{0110}$, $\ket{0111}$, $\ket{1000}$, $\ket{1001}$, $\ket{1010}$, $\ket{1011}$, $\ket{1100}$, $\ket{1101}$, $\ket{1110}$, $\ket{1111}\}$. In this basis, $0$ indicates the presence of an electron, while $1$ represents a vacuum (no charge) in the corresponding quantum dot. This notation is particularly well-suited for quantum computing procedures. Moreover, the $2^4$-dimensional space is ideal for treating the system as an open system, where the total number of particles may fluctuate as electrons flow between the reservoirs and the qubits.

\subsection{Eigenvalues}

With no intra-qubit hopping ($\gamma=0$), the Hamiltonian given by Eq. (\ref{H0}) becomes diagonal in the computational basis described above. The corresponding set of eigenenergies is presented in Table \ref{tab1}. The zero energy corresponds to states in which at least one of the qubits is empty. The energy $J'$ arises when both qubits are singly occupied, with the left (right) dot of the upper qubit and the right (left) dot of the lower qubit being occupied. The energy $J$ occurs when both qubits are singly occupied, with the left (right) dot of the upper qubit and the left (right) dot of the lower qubit being occupied. The energy $J+J'$ appears when one of the qubits is doubly occupied while the other remains singly occupied. Lastly, the energy $2J+2J'$ is observed when both qubits are doubly occupied.

\begin{table}[h]
\centering
\begin{tabular}{|c|c|}
\hline
eigenenergy ($\varepsilon_i$) & eigenstates \\ \hline
0    & $\ket{0011}, \ket{0111}, \ket{1011}, $   \\ 
     &  $\ket{1100},\ket{1101}, \ket{1110},$    \\
     &  $ \ket{1111}$                           \\ \hline
$J'$    & $\ket{0110}, \ket{1001}$              \\ \hline
$J$    & $\ket{0101}, \ket{1010}$               \\ \hline
$J + J'$    & $\ket{0001}, \ket{0010}, \ket{0100},$   \\ 
     & $\ket{1000}$                                  \\ \hline
$2J + 2J'$    & $\ket{0000}$                           \\ \hline
\end{tabular}
\caption{Eigenenergies and corresponding eigenstates for $\gamma=0$ (no intra-qubit hopping). 
The quantities $J$ and $J'$ gives the direct and indirect Coulomb interaction between the qubits.
\label{tab1}}
\end{table}

\begin{figure}
	\includegraphics[scale=0.3]{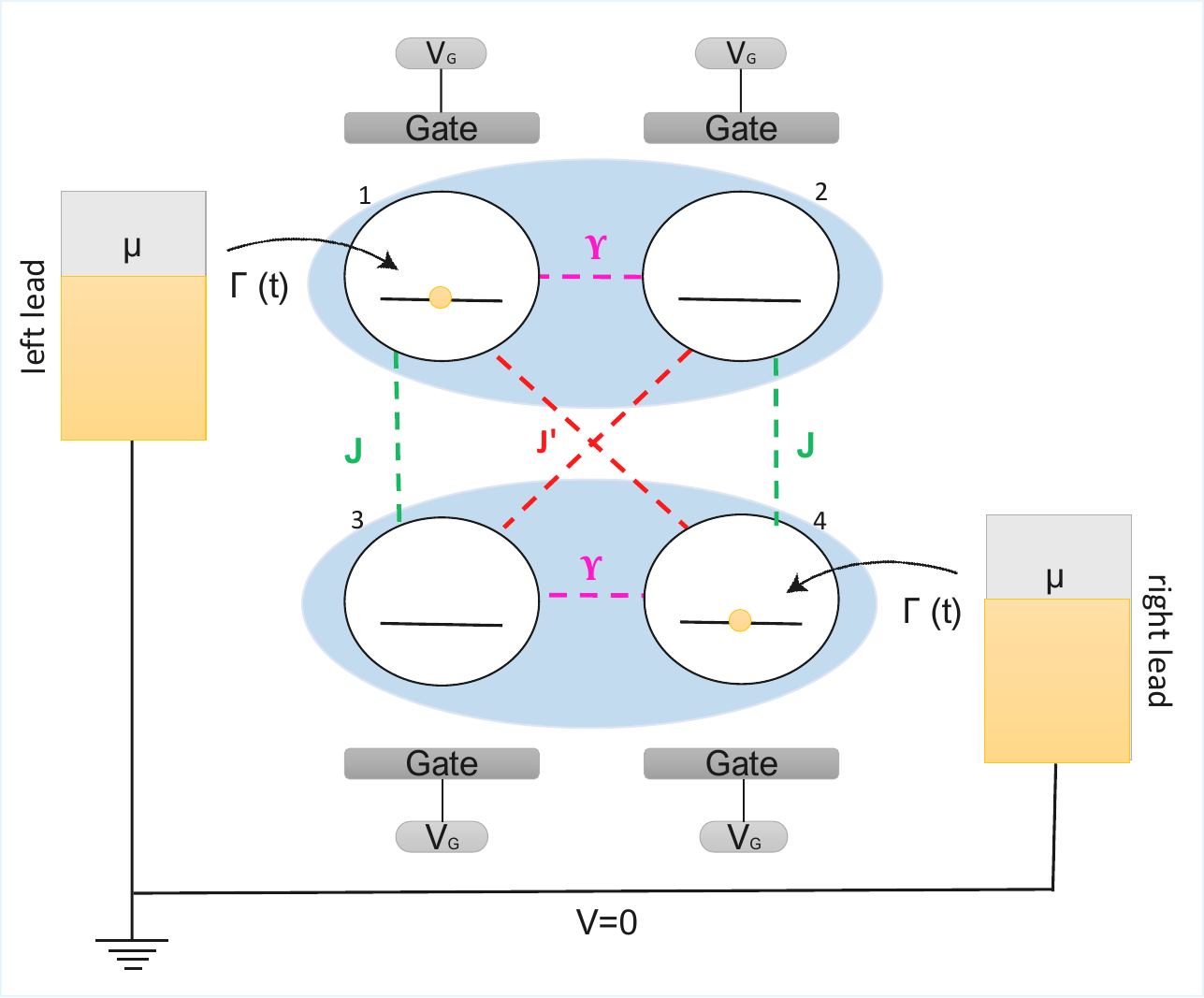}
	\caption{\label{system}Illustration of the system of interest. Four quantum dots labeled as 1, 2, 3, 4 are arranged in an array that forms a bipartite two-qubit structure. 
	The upper dots 1 and 2 hybridizes their orbitals thus forming a molecular structure that provides one charge qubit. Similarly, the lower dots 3 and 4 constitutes a second qubit. No charge can flow between the qubits, however they are capacitivelly coupled to each other via Coulomb interactions with strength $J$ and $J'$. Left and right leads can inject charge into the qubits. }
\end{figure}

\subsection{Unitary Quantum Evolution: Analytical Approach to Closed Qubits}

In the following, we examine the subspace defined by singly occupied qubit states $\{\ket{0110}, \ket{1001}\}$. By appropriately adjusting $J$ and $J'$, with $J > J'$, the energy of these states can be significantly separated from other levels, thereby creating more favorable conditions for the formation of entangled states, such as:
\begin{align}
\ket{\phi} =\frac{\ket{0110}+e^{i\phi}\ket{1001}}{\sqrt{2}} \label{target},
\end{align}
which will serve as our target state in the calculation presented below. In our physical system, the state $\ket{0110}$ is not directly coupled to $\ket{1001}$ for $\gamma \neq 0$. However, higher-order processes enable a transition between these two states. By employing second-order perturbation theory, the effective coupling parameter $\Omega$ can be determined as:
\begin{eqnarray}\label{Omega} \Omega = \sum_{i} \frac{ \bra{0110} v \ket{i} \bra{i} v \ket{1001}}{\varepsilon - \varepsilon_{i}} = - \frac{2\gamma^2}{(J-J')}, 
\end{eqnarray}
where the summation is performed over all possible intermediate transition states.With this effective coupling, the time evolution of the system within the two-level subspace $\{\ket{0110}, \ket{1001}\}$ is given by:
\begin{align}
	\label{psit} \ket{\psi(t)} = \cos\left(\frac{\lvert \Omega \rvert t}{\hbar}\right)\ket{0110} + i\sin\left(\frac{\lvert \Omega \rvert t}{\hbar}\right)\ket{1001}, 
\end{align}
where $\ket{0110}$ is considered the initial state. The period of the dynamics is expressed as:
\begin{equation} 
	T = \frac{2 \pi \hbar}{|\Omega|} = \frac{\pi \hbar}{\gamma^2} (J-J'), 
\end{equation}
and the first maximally entangled state occurs at time $\tau$, defined as:
\begin{equation}
	\label{tmax} 
	\tau = \frac{\pi \hbar}{4 |\Omega|} = \frac{\pi \hbar}{8 \gamma^2} (J-J') = \frac{T}{8}. 
\end{equation}
Thus, the entanglement formation time can be tuned by adjusting $\gamma$ and the energy difference $J-J'$. To simplify our analysis, we assume $J = 2J'$, which leads to $\tau = \pi \hbar J' / (8\gamma^2)$ and $T = \pi \hbar J' / \gamma^2$. By parameterizing $\gamma$ in terms of $J'$, i.e., $\gamma = x J'$, we can express $\tau$ and $T$ as $ \pi \hbar / (8 J' x^2)$ and $ \pi \hbar / (J' x^2)$, respectively. 

In our model, we consider physically feasible values of $J' = 200$ $\mu eV$ and $x = 0.05$, which correspond to $\gamma = 10$ $\mu eV$~\cite{fujisawa2011} so $\tau \approx 0.5$ ns and $T \approx 4$ ns.  When comparing this time scale to the phase flip errors typically observed in semiconductor nanostructures, characterized by a dephasing time of $\tau_{deph} = 1.0$ ns (corresponding to a 1 GHz rate), it becomes evident that dephasing can significantly constrain the ability to achieve high levels of entanglement within this window. This highlights the importance of incorporating dephasing effects into our model to accurately capture the dynamics of the system and its limitations. Decoherence rates can also be estimated following a procedure via transport analysis~\cite{valente10,vorojtsov05}.

\subsection{Coupling to the Environment}
Initialization is inherently an incoherent process, as it involves interaction between the system and particle sources and drains. This coupling introduces additional channels of decoherence that can perturb the system and leave lingering effects on entanglement formation, even after the gate pulse is terminated. To model the initialization dynamics, we employ the Lindblad equation formulated in terms of second-quantized operators.
The quantum dots coupled to fermionic reservoirs can be described as
\begin{equation}\label{V}
V (t) = \sum_{\mathbf{k}, n} [V_{\mathbf{k},n}^*(t) d_n^\dagger c_{\mathbf{k},n} + V_{\mathbf{k},n} (t) c_{\mathbf{k},n}^\dagger d_n  ],
\end{equation}
where $c_{\mathbf{k},n}$ ($c_{\mathbf{k},n}^\dagger$) annihilates (creates) an electron with wave vector $\mathbf{k}$ in the $n$-th reservoir, the matrix element
$V_{\mathbf{k},n} (t)$ defines a time-dependent coupling parameter that governs the interaction between state $\mathbf{k}$ in the lead $n$ and quantum dot $n$.
So the full model is given by
\begin{equation}
H(t) = H_0 + H_R + V(t),
\end{equation}
where $H_0$ is given by Eq. (\ref{H0}) and $H_R$ is the reservoir Hamiltonian (free particle). This complete model will be applied to derive the Lindblad equation.

\subsection{Lindblad Equation}

Consider the von Neumann equation $\dot{\rho}(t) = -i [H(t), \rho(t)]$. Moving to the interaction picture, the equation becomes
$\dot{\hat{\rho}}(t) = -i [\hat{V}(t), \hat{\rho}(t)]$ which, after integration, we can write as
\begin{eqnarray}
\dot{\hat{\rho}}(t) = -i [\hat{V}(t), \hat{\rho}(0)] - \int_0^t dt_1 [\hat{V}(t),[\hat{V}(t_1),\hat{\rho}(t_1)]], \phantom{x}
\end{eqnarray}
resulting in the four terms below
\begin{eqnarray}
&& \dot{\hat{\rho}}(t) = -i [\hat{V}(t), \hat{\rho}(0)] - \int_0^t dt_1 \{  \hat{V}(t) \hat{V}(t_1) \hat{\rho}(t_1) - \\
&& \phantom{xx} \hat{V}(t) \hat{\rho}(t_1) \hat{V}(t_1)  - \hat{V}(t_1) \hat{\rho}(t) \hat{V}(t) +  \hat{\rho}(t_1) \hat{V}(t_1) \hat{V}(t)\} . \nonumber
\end{eqnarray}
Considering the Born-Markov approximation and taking the partial trace over the electronic reservoirs (R), we can write
\begin{eqnarray}\label{rhoQ_1}
&& \dot{\hat{\rho}}_Q(t) = - \int_0^t dt_1 \mathrm{Tr}_{R}\{  \hat{V}(t) \hat{V}(t_1) \hat{\rho}_Q(t_1) \otimes \rho_{R} - \nonumber \\
&& \phantom{xx} \hat{V}(t)  \hat{\rho}_Q(t_1) \otimes \rho_{R} \hat{V}(t_1)  - \hat{V}(t_1)  \hat{\rho}_Q(t_1) \otimes \rho_{R} \hat{V}(t) + \nonumber \\
&&  \phantom{xx} \hat{\rho}_Q(t_1) \otimes \rho_{R}\hat{V}(t_1) \hat{V}(t)\} ,
\end{eqnarray}
where $ \rho_{R} = \rho_R(0)$ is the time-independent density matrix for the reservoirs and $\hat{\rho}_Q(t)$ is the qubits subsystem densidy matrix. 
Using Eq. (\ref{V}) we arrive at
\begin{eqnarray}\label{rhoQ_3}
&&\dot{\hat{\rho}}_Q(t) = - \int_0^t dt_1  \sum_{\mathbf{k}, n} \{ \nonumber \\ 
&& +V_{\mathbf{k},n}^*(t) V_{\mathbf{k},n}(t_1) \hat{d}_n^\dagger(t) \hat{d}_n(t_1) \hat{\rho}_Q(t_1) \langle \hat{c}_{\mathbf{k},n}(t)  \hat{c}_{\mathbf{k},n}^\dagger(t_1) \rangle \nonumber \\
&& +V_{\mathbf{k},n}(t) V_{\mathbf{k},n}^*(t_1) \hat{d}_n(t) \hat{d}_n^\dagger (t_1) \hat{\rho}_Q(t_1) \langle  \hat{c}_{\mathbf{k},n}^\dagger(t)  \hat{c}_{\mathbf{k},n}(t_1) \rangle \nonumber \\
&& - V_{\mathbf{k},n}^*(t) V_{\mathbf{k},n}(t_1) \hat{d}_n^\dagger(t) \hat{\rho}_Q(t_1) \hat{d}_n(t_1)  \langle \hat{c}_{\mathbf{k},n}^\dagger(t_1) \hat{c}_{\mathbf{k},n}(t)   \rangle \nonumber \\
&& - V_{\mathbf{k},n}(t) V_{\mathbf{k},n}^*(t_1) \hat{d}_n(t) \hat{\rho}_Q(t_1) \hat{d}_n^\dagger (t_1)  \langle \hat{c}_{\mathbf{k},n}(t_1) \hat{c}_{\mathbf{k},n}^\dagger(t)   \rangle \nonumber \\
&& - V_{\mathbf{k},n}^*(t_1) V_{\mathbf{k},n}(t)\hat{d}_n^\dagger(t_1) \hat{\rho}_Q(t_1) \hat{d}_n(t)  \langle \hat{c}_{\mathbf{k},n}^\dagger(t) \hat{c}_{\mathbf{k},n}(t_1)   \rangle \nonumber \\
&& - V_{\mathbf{k},n}(t_1) V_{\mathbf{k},n}^*(t) \hat{d}_n(t_1) \hat{\rho}_Q(t_1) \hat{d}_n^\dagger (t)  \langle \hat{c}_{\mathbf{k},n}(t) \hat{c}_{\mathbf{k},n}^\dagger(t_1)   \rangle \nonumber \\
&& +V_{\mathbf{k},n}^*(t_1) V_{\mathbf{k},n}(t) \hat{\rho}_Q(t_1) \hat{d}_n^\dagger(t_1) \hat{d}_n(t) \langle \hat{c}_{\mathbf{k},n}(t_1)  \hat{c}_{\mathbf{k},n}^\dagger(t) \rangle \nonumber \\
&& +V_{\mathbf{k},n}(t_1) V_{\mathbf{k},n}^*(t) \hat{\rho}_Q(t_1) \hat{d}_n(t_1) \hat{d}_n^\dagger (t) \langle  \hat{c}_{\mathbf{k},n}^\dagger(t_1)  \hat{c}_{\mathbf{k},n}(t) \rangle \},\nonumber \\
\end{eqnarray}
where $\langle \hat{c}_{\mathbf{k},n}(t)  \hat{c}_{\mathbf{k},n}^\dagger(t_1) \rangle = \mathrm{Tr}_{R} \{ \hat{c}_{\mathbf{k},n}(t)  \hat{c}_{\mathbf{k},n}^\dagger(t_1) \rho_R(0) \}$
and similar definitions to the other averages. Notice that
\begin{eqnarray}
\langle \hat{c}_{\mathbf{k},n}(t)  \hat{c}_{\mathbf{k},n}^\dagger(t_1) \rangle &=& (1-f_{\mathbf{k},n}) e^{-i \xi_{\mathbf{k},n} (t-t_1)}, \\
\langle  \hat{c}_{\mathbf{k},n}^\dagger(t)  \hat{c}_{\mathbf{k},n}(t_1) \rangle &=& f_{\mathbf{k},n} e^{-i \xi_{\mathbf{k},n} (t_1-t)}, \\
\langle \hat{c}_{\mathbf{k},n}^\dagger(t_1) \hat{c}_{\mathbf{k},n}(t)   \rangle &=& f_{\mathbf{k},n} e^{-i \xi_{\mathbf{k},n} (t-t_1)}, \\
\langle \hat{c}_{\mathbf{k},n}(t_1) \hat{c}_{\mathbf{k},n}^\dagger(t)   \rangle &=& (1-f_{\mathbf{k},n}) e^{-i \xi_{\mathbf{k},n} (t_1-t)},
\end{eqnarray}
here, $f_{\mathbf{k},n}$ denotes the Fermi function for reservoir $n$, and $\xi_{\mathbf{k},n}$ represents the corresponding free-electron energy. 
We assume that the chemical potential in the source leads is sufficiently high such that $f_{\mathbf{k},n} = 1$, and that $V_{\mathbf{k},n}(t)=V_n(t)$ is time-dependent but independent of $\mathbf{k}$.
By replacing the summation over wavevectors $\sum_{\mathbf{k}}$ with an integral over energy, we obtain the correspondence $\sum_{\mathbf{k}} \to \int d\xi\, D_n$, 
where $D_n$ denotes the density of states of reservoir $n$. Assuming $D_n$ to be energy-independent, Eq. (\ref{rhoQ_3}) can be rewritten as
\begin{eqnarray}\label{rhoQ_4}
&&  \dot{\hat{\rho}}_Q(t) = - \frac{1}{2} \sum_{ n}  \Gamma_n(t) \{  \hat{d}_n(t) \hat{d}_n^\dagger (t) \hat{\rho}_Q(t) \nonumber \\ 
&& - 2 \hat{d}_n^\dagger(t) \hat{\rho}_Q(t) \hat{d}_n(t) + \hat{\rho}_Q(t) \hat{d}_n(t) \hat{d}_n^\dagger (t)  \},
\end{eqnarray}
with $ \Gamma_n(t)=2\pi D_n |V_n(t)|^2$ being a tunneling rate between a quantum dot and its corresponding source lead.
Expressing Eq. (\ref{rhoQ_4}) in the Schr\"odinger picture we arrive at
\begin{align}
&\dot{\rho_Q}(t) = -\frac{i}{\hbar}[H_0, \rho_Q(t)] + \nonumber \\
    &\frac{1}{2}\frac{\Gamma(t)}{\hbar} \sum_{n}  ( 2 d_n^\dagger \rho_{Q}(t) d_n -d_n d_n^\dagger \rho_Q(t) - \rho_Q(t) d_n d_n^\dagger ),\label{lindblad}
\end{align}
where $\hbar$ has been reintroduced. It is also convenient to write Eq. (\ref{lindblad}) in terms of the unitless variable $\theta=|\Omega| t / \hbar$, as
\begin{align}
 &\frac{\partial \rho_Q(\theta)}{\partial \theta}= -i[\frac{H_T}{|\Omega|}, \rho_S(\theta)] + \nonumber \\
    &\frac{1}{2}\frac{\Gamma(\theta)}{|\Omega|} \sum_{n}  ( 2 d_n^\dagger \rho_{S}(\theta) d_n -d_n d_n^\dagger \rho_S(\theta) - \rho_S(\theta) d_n d_n^\dagger ).\label{lindblad2}
\end{align}

\subsection{Gate Modulated Tunneling Rate}

The behavior of the system during the pulse is pivotal to its evolution and the formation of highly entangled states. In the following, we examine two types of pulses that modulate the rate 
$\Gamma(t)$, i.e., (\emph{i}) a square pulse and (\emph{ii}) a Gaussian pulse. In case (\emph{i}), we assume that $\Gamma(t) = \Gamma_0 + \delta(t)$ for $0 < t < \sigma$, and $\Gamma(t) = \delta(t)$ otherwise.
Here, $\sigma$ denotes the pulse width. Furthermore, we define $\sigma$ as a fraction $p$ of the period $T$, such that $\sigma = pT$. The parameter $\Gamma_0$ represents a time-independent tunneling coupling strength. 
The term $\delta(t)$ accounts for stochastic variations in the gate voltages, introduced to mimic the electric amplitude noise present in the system.
In terms of the variable $\theta$, the pulse width can be defined as $\sigma_\theta = |\Omega| \sigma / \hbar$, or more simply as $\sigma_\theta = 2 \pi p$. It is worth noting that charge transport in nanostructures under a rectangular bias pulse was originally studied by Wingreen \textit{et al.}~\cite{wingreen1993} and Jauho \textit{et al.}~\cite{jauho1994} using Keldysh nonequilibrium Green functions. In this work, we adapt these concepts for quantum computation to initialize two charge qubits. To simulate a Gaussian pulse in case (\emph{ii}) we adopt the profile $\Gamma(\theta) = \Gamma_0 e^{-[(\theta-\theta_0)/\sigma_\theta]^2} +  \delta(\theta)$, where $\theta_0$ denotes the pulse center. This specific type of pulse has also been investigated in theoretical studies that consider an alternative qubit encoding in coupled quantum dots~\cite{platero1}. In our work, we focus exclusively on these two types, as charge qubits cannot be implemented using mechanisms based on spin-orbit interactions, which are available only to spin qubits~\cite{platero2}.
We assume sub-nanoseconds voltage pulses in our treatment. The ultrafast square gate pulses are experimentally accessible through state-of-the-art electronics\cite{wehbi2020, wehbi2022, zhau2014}. Because of the short timescale, our model also includes the effect of amplitude noise, as it will be discussed in the next section.

\section{Results}
\label{sec:results}
In this section, we present our findings on the system initialization aimed at optimizing the creation of a highly entangled state. 
We first assume that the two qubits are initially empty of charge, with their state described by $\ket{1111}$ ($\bra{1111} n_i \ket{1111} = 0$ for $i = 1,...,4$). To evolve the quantum system in Eq. (\ref{target}), the transition $\ket{1111}$ to $\ket{0110}$ must be performed. This transition requires populating quantum dots QD1 and QD4. The process can be achieved by injecting electrons from the source leads into QD1 and QD4 in a time-controlled manner, as illustrated in Fig.~\ref{system}. However, when a gate voltage pulse such as the square (\emph{i})  and gaussian (\emph{ii}) is applied, the system does not exclusively reach the desired state $\ket{0110}$. For instance, states such as $\ket{0000}$, $\ket{1000}$, and others may have a finite probability of being occupied after initialization, resulting in poorly entangled states. 
\begin{figure}
	\includegraphics[scale=0.5]{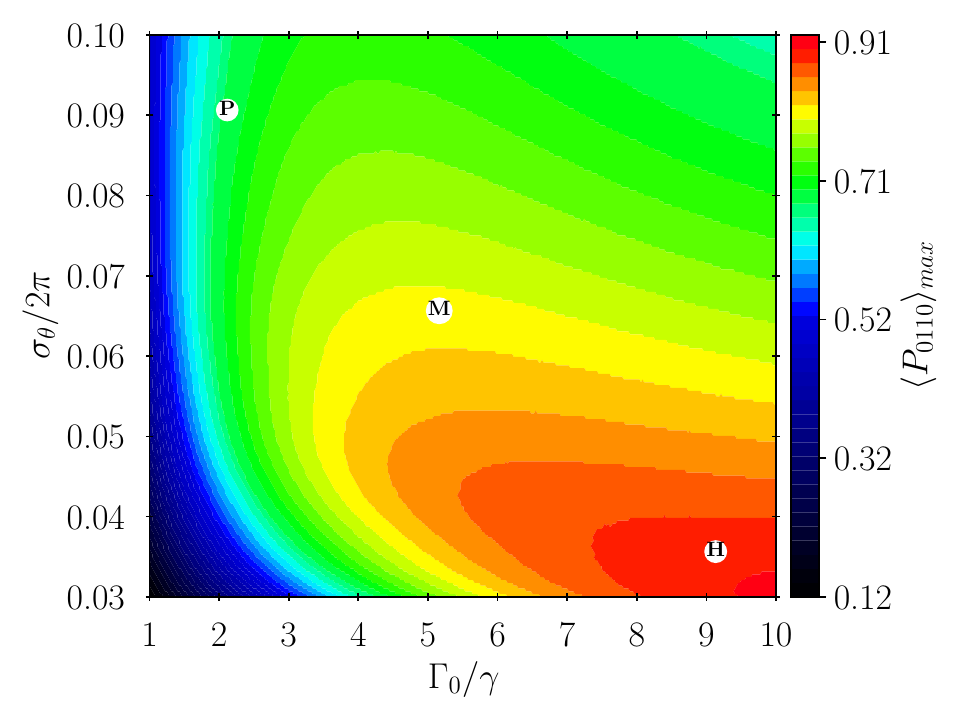}
	\caption{The maximum fidelity for the initial target state $\ket{0110}$ as a function of the electronic pulse parameters: intensity ($\Gamma_0$) and width ($\sigma_\theta$). Small values of $\sigma_\theta$ combined with large values of $\Gamma_0$ provide optimal initialization conditions. The fidelity does not reach 100\%, as additional states may also be populated during the initialization pulse. In the figure, three specific points are highlighted with the letters H, M, and P, corresponding to high, medium, and poor-quality initialization conditions, respectively.\label{fig2}}
\end{figure}

In Fig.~\ref{fig2}, we show the maximum value of the fidelity to the state $P_{0110} = \ket{0110}\bra{0110}$ achieved after the initialization of the square pulse (\emph{i}) with no noise ($\delta=0$), i.e., $\langle P_{0110}\rangle_{\mathrm{max}}=\mathrm{Max}\left\{{\mathrm{Tr}[\rho(t) P_{0110}]}\right\}$, as a function of both, the leads-dots tunneling rate $\Gamma_0$ and the pulse width $\sigma_\theta/2\pi$. From this plot, we observe a range of optimized parameters, with $\sigma_\theta/2\pi$ around $0.03$ to $0.06$ ($3\%$ to $6\%$ of the period $T$, respectively) and $\Gamma_0$ between $5$ to $10$ times the intra-qubit hopping parameter $\gamma$. As $\Gamma_0$ decreases, the charge injection becomes too weak to properly initialize the system in the state $\ket{0110}$. Conversely, if pulse width $\sigma_\theta/2\pi$ becomes too large, around $0.07$, the initialization is compromised due to prolonged exposure of the qubits to the leads, which starts to activate additional states of the system and introduces additional decoherence sources. 

We also found that higher fidelity to the state $\ket{0110}$ after pulse initialization leads to better entanglement formation. We quantify the entanglement degree through the negativity, $\mathcal{N}$, obtained by calculating
\begin{equation}
\mathcal{N}=\frac{1}{2}\sum\limits_{k} \left(|\mathcal{L}_{k}|-\mathcal{L}_{k}\right),
\label{eqn:Negativity}
\end{equation}
which is based on the PPT criterion for separability~\cite{Horodecki1996a, Horodecki1996b,peres1996, Horodecki2009}. This quantity measures the entanglement degree of the evolved state. Here $\lbrace\mathcal{L}_{k}\rbrace$ is the $k$-th  eigenvalue of $\rho^{T_{1}}$, which is the partial transpose of the density matrix $\rho$ of the full system concerning the subsystem corresponding to the first qubit (QD1-QD2). Equivalently, it can be defined in terms of the partial transpose $\rho^{T_{2}}$ with transposition being taken to the second qubit (QD3-QD4). 

\begin{figure}
	\includegraphics[scale=0.5]{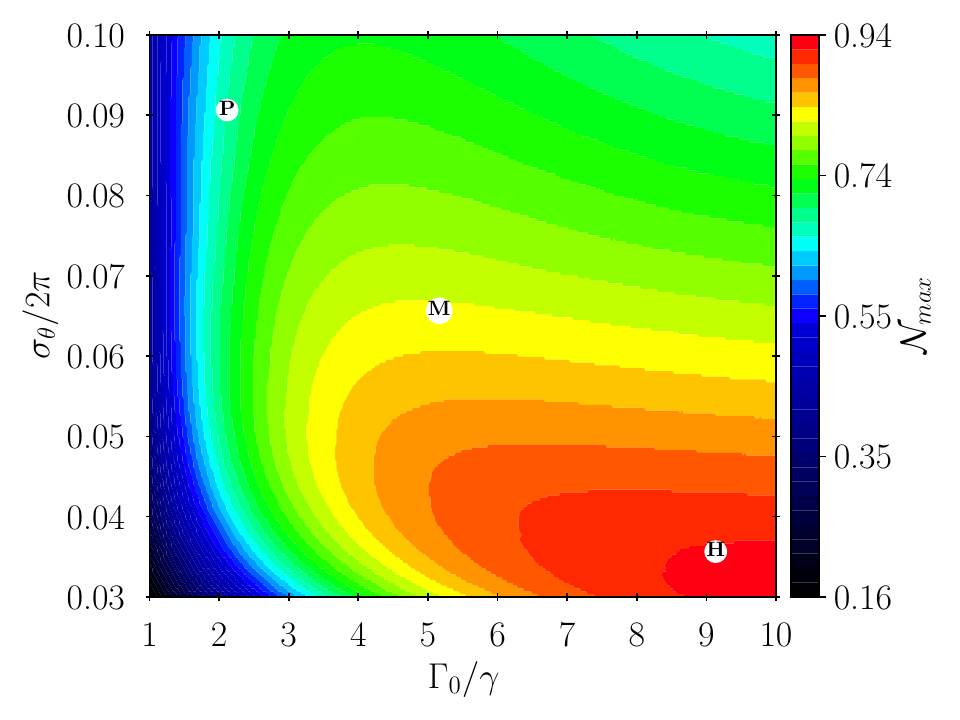}
	\caption{Maximum value of negativity, $\mathcal{N}$, after pulse initialization as a function of $\Gamma_0$ and $\sigma_\theta$. The negativity behavior shows a strong correlation with the fidelity in Fig.~\ref{fig2}, indicating that achieving high-quality initialization is essential for forming a highly entangled state. The same points H, M and P, from Fig.~\ref{fig2}, are highlighted here.\label{fig3}}
\end{figure}

Our results for the maximum value of the negativity obtained in the temporal evolution, considering the same range of physical parameters as in Fig.~\ref{fig2}, are shown in Fig.~\ref{fig3}. Note that the plot exhibits behavior similar to that found for the fidelity to the state $\ket{0110}$, as seen in Fig.~\ref{fig2}. The negativity reaches high values for faster pulses and larger tunnel injection of carriers. However, we must consider possible technological limitations regarding time-dependent gate voltages. If we set the pulse width $\sigma_{\theta}$ as short as $0.05$ ($5\%$ of the dynamical period $T$), we find $\sigma \approx 0.2$ ns. Therefore, it is not simply a matter of reducing the pulse width: we need to adjust both $\sigma_{\theta}$ and $\Gamma_0$ to optimize the initialization while maintaining experimental feasibility~\cite{wehbi2020, wehbi2022, zhau2014}.
\begin{figure}
        \centering
	\includegraphics[scale=0.55]{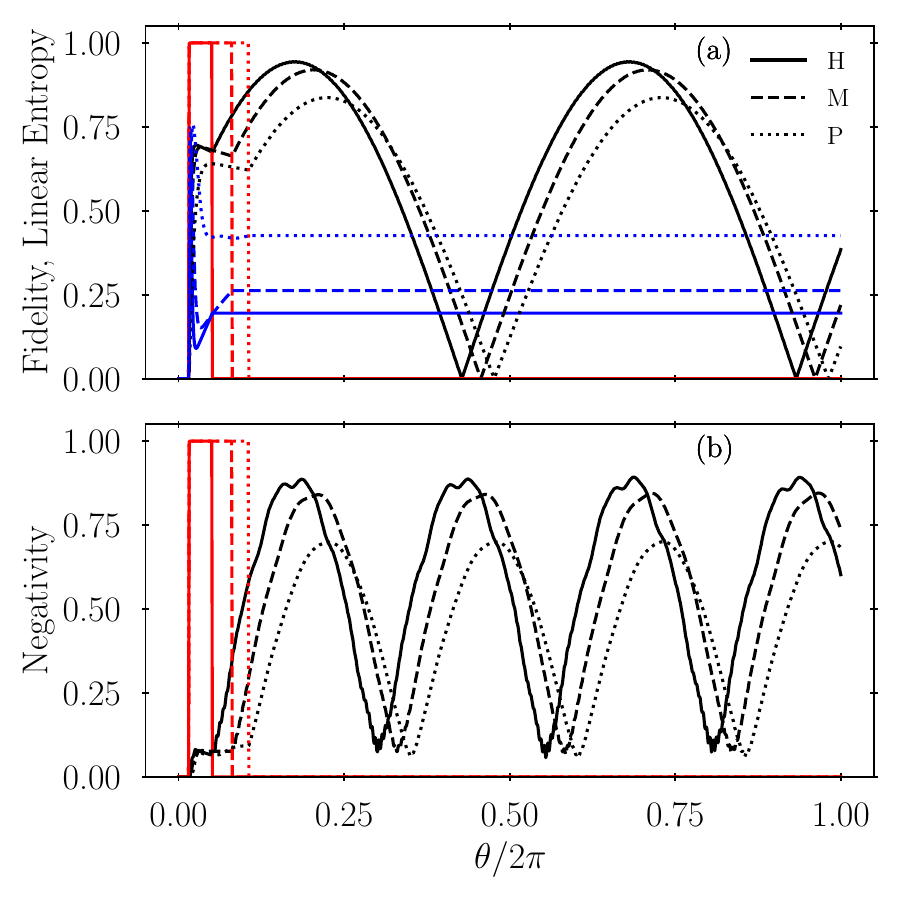}
	\caption{(Color online) Panel (a): Fidelity (black lines) and linear entropy (blue lines) as a function of $\theta/2\pi$ for the three points in Figs.~\ref{fig2}-\ref{fig3}, corresponding to high (H) (solid line), medium (M) (dashed line), and poor (P) (dotted line) initialization quality. The rectangular pulse (red lines) is illustrated for the three cases. The fidelity $\mathcal{F}$ is calculated for $\sigma_{tar}=\ket{\phi}\bra{\phi}$, with $\phi=\pi/2$.  Panel (b): Negativity $\mathcal{N}$ against $\theta$ for the three cases H, M, and P. Note that both fidelity $\mathcal{F}$ and negativity $\mathcal{N}$ attain higher values in the H case and become more suppressed in the M and P cases. Observe that $\mathcal{N}$ also peaks at the fidelity dips, indicating that the evolved state reaches highly entangled states twice within a fidelity period.
	 \label{fig4}}
\end{figure}

Based on Fig.~\ref{fig3}, we now proceed to analyze the quantum dynamics after initialization, considering the highlighted points in the later figures: high (H), middle (M), and poor (P) quality initialization conditions.The applied rectangular pulse is depicted by red lines in all three cases.  The parameters $(\Gamma_0/\gamma, \sigma_\theta/2\pi)$ for H, M, and P are $(9, 0.035)$, $(5, 0.065)$, and $(2, 0.09)$, respectively. In Fig.~\ref{fig4}(a), we show the fidelity $\mathcal{F}$ as function of time, parameterized by $\theta$, for the three cases considered. The fidelity is measured against the target state in Eq.~(\ref{target})~\cite{nielsenchuang2007},
\begin{equation}
 \mathcal{F}(\rho_S, \sigma)=\mathrm{Tr}\{\sqrt{\sqrt{\rho_S}\sigma\sqrt{\rho_S}}\},
\end{equation}
where $\sigma=\ket{\phi}\bra{\phi}$ with $\phi=\pi/2$.
For simplicity, using Eq.~(\ref{target}) and Eq.~(\ref{psit}), we derive an analytical expression for the fidelity, valid for a closed system initialized with 100\% accuracy in the state $\ket{0110}$. This results in:
\begin{equation}\label{fanalitic}
 \mathcal{F}_{\mathrm{analitic}}(\ket{\psi(t)}\bra{\psi(t)}, \sigma) =  \sqrt{ [1+\sin(2\theta)]/2},
\end{equation}
which attains maximum value at $\theta/2\pi=1/8=0.125$, $5/8$, etc.

In Fig.~\ref{fig4}(a), the fidelity $\mathcal{F}$ reaches values close to 0.9 in the H choice, revealing that the system approaches the highly entangled state described by Eq.~(\ref{target}) with $\phi=\pi/2$. In the M and P cases, the fidelity decreases, indicating a lesser approach to the target state. Notice that the positions of the maxima closely follow the conditions for $\mathcal{F}_{\mathrm{analitic}}=1.0$ in the analytical Eq.~(\ref{fanalitic}). Nevertheless, even in the best scenario (H), the fidelity does not reach this value. The explanation is mainly due to two factors: (i) The initialization process does not generate solely the initial desired state $\ket{0110}$, as undesired states can also be populated. As seen in Fig.~\ref{fig2}, the maximum fidelity with respect to the state $\ket{0110}$ is around 90\% after the square pulse. (ii) The initialization based on carrier injection from source leads is an incoherent process. Even in the absence of internal dephasing mechanisms, the quantum dynamics become somewhat messy during initialization.

We also show Fig.~\ref{fig4}(a) in the linear entropy $\mathcal{S}=1-\mathrm{Tr}(\rho_S^2)$ of the physical system tracing the reservoirs as blue lines. In all three cases (H, M, and P) the entropy $\mathcal{S}$ attains non-zero values after the pulse. This indicates that the initialization does not bring the system to a fully pure state, even for the best set of parameters H, which results in $\mathcal{S}\approx 0.2$. Nevertheless, the linear entropy $\mathcal{S}$ attains lower stationary values after the initialization pulse in the H regime compared to those found in the M and P cases.

The fidelity shown in Fig.~\ref{fig4}(a) demonstrates that the maximum entangled state in Eq.~(\ref{target}) with $\phi=\pi/2$ can be closely achieved. However, this does not exclude the possibility that other entangled states are also being formed during the dynamics. To quantify the entanglement, Fig.~\ref{fig4}(b) presents our results for the evolution of the negativity $\mathcal{N}$ under the same setups. We observe that the maxima appear not only at the times where the fidelity attains its maxima but also at the dips. This indicates that the fidelity dips correspond to a highly entangled state, as described by Eq.~(\ref{target}), but with a relative phase $\phi$ rotated by $\pi$. Consequently, the number of entangled states in an oscillation period is twice the number of fidelity peaks.

\begin{figure}
    \centering
	\includegraphics[scale=0.53]{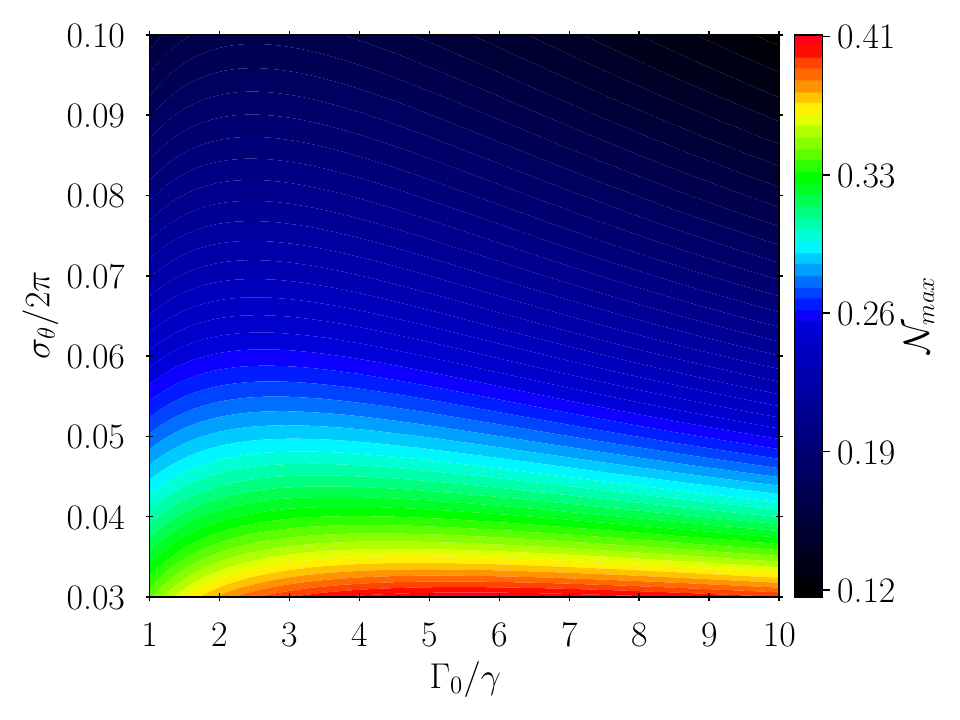}
	\caption{Maximum value of negativity, $\mathcal{N}$, after pulse initialization as a function of $\Gamma_0$ and $\sigma_\theta$ for a Gaussian pulse. Compared to Fig. (\ref{fig3}), which employed a square pulse yielding $\mathcal{N} \approx 0.9$, the Gaussian pulse here results in markedly lower entanglement, peaking near $0.4$. This highlights the decisive influence of pulse shape on entanglement generation. \label{fig7}}
\end{figure}

We now extends this study by examining entanglement formation using a Gaussian-shaped gate pulse. As shown in Fig. (\ref{fig7}), we report markedly lower values of entanglement, quantified via negativity, following electronic initialization with the Gaussian profile. Specifically, the maximum negativity observed is around 0.4, in contrast to $\mathcal{N}\approx 0.9$ achieved with square pulses found in Fig. (\ref{fig3}). The results above highlights an often-overlooked aspect of hardware control, the pulse profile, as a key parameter in entanglement formation. Our analysis suggest that even at the initialization stage, the shape of gate pulse can significantly constrain or enhance the quantum correlations subsequently formed and emphasizes that pulse engineering is not merely a technical detail, but a foundational factor in entanglement dynamics. Furthermore, the flexibility of our theoretical framework, which applies Eq. (\ref{lindblad}) based on second-quantized operators for electronic devices, readily accommodates other pulse profiles. This opens possibilities to systematic studies of pulse-shape-dependent initialization schemes, potentially optimizing entanglement for diverse quantum information protocols.

\begin{figure}
    \centering
	\includegraphics[scale=0.53]{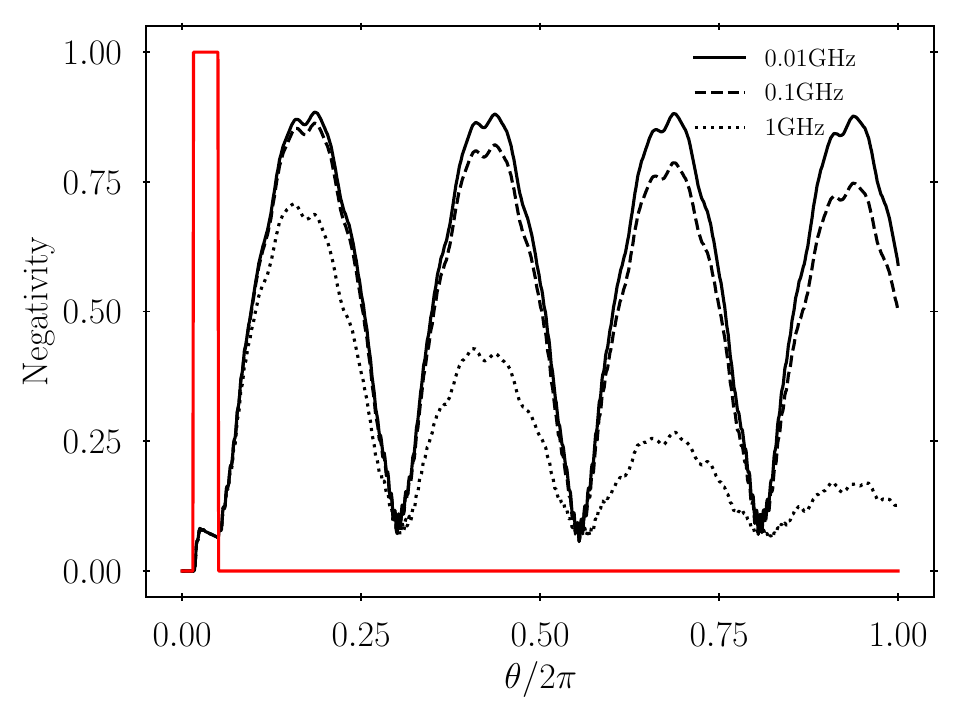}
	\caption{(Color online) Negativity $\mathcal{N}$ in the H regime is shown for three dephasing rates: $10^{-2}$ GHz (solid line), $10^{-1}$ GHz (dashed line), and 1 GHz (dotted line). The applied pulse is also illustrated (solid red line). For lower dephasing rates, negativity persists throughout the first cycle, reaching values close to 0.8. As the dephasing increases, negativity decays more rapidly, indicating a suppression of entanglement generation.
	\label{fig5}}
\end{figure}

Finally, to mimic dephasing sources imposed by noise sources such as phase flip errors, we add one more collapse operator to the Lindbladian,
\begin{equation}\label{lindblad_dephasing}
  \mathcal{L}_2( \rho_Q) = \frac{1}{2} \sum_{n=1}^2  ( 2 C_n \rho_{S}(\theta) C_n^\dagger - C_n^\dagger C_n \rho_S(\theta) - \rho_S(\theta) C_n^\dagger C_n ),
\end{equation}
where $C_n= \sqrt{\Gamma_{dph}/\Omega} P_n$, with $P_1=\ket{0110}\bra{0110}$ and $P_2=\ket{1001}\bra{1001}$. In Fig.~\ref{fig5}, we show the evolution of negativity for three distinct phase flip error ratios: $\Gamma_{dph}=10^{-2}$ GHz, $10^{-1}$ GHz, and $1$ GHz. The values were selected as fractions of a realistic benchmark~\cite{fujisawa2011}, to highlight the potential advantages of directing experimental efforts toward reducing the decoherence time in this specific system.  The pulse parameter is taken to the best-case scenario found for the square pulse, point H in Fig. (\ref{fig3}). For $10^{-2}$ GHz, the negativity remains high, similar to the case with no dephasing. For $10^{-1}$ GHz, it is slightly suppressed, while for $1$ GHz, a significant suppression is observed. Therefore, future experimental implementations of the present double charge qubits must be able to mitigate possible dephasing sources.

\begin{figure}
    \centering
	\includegraphics[scale=0.45]{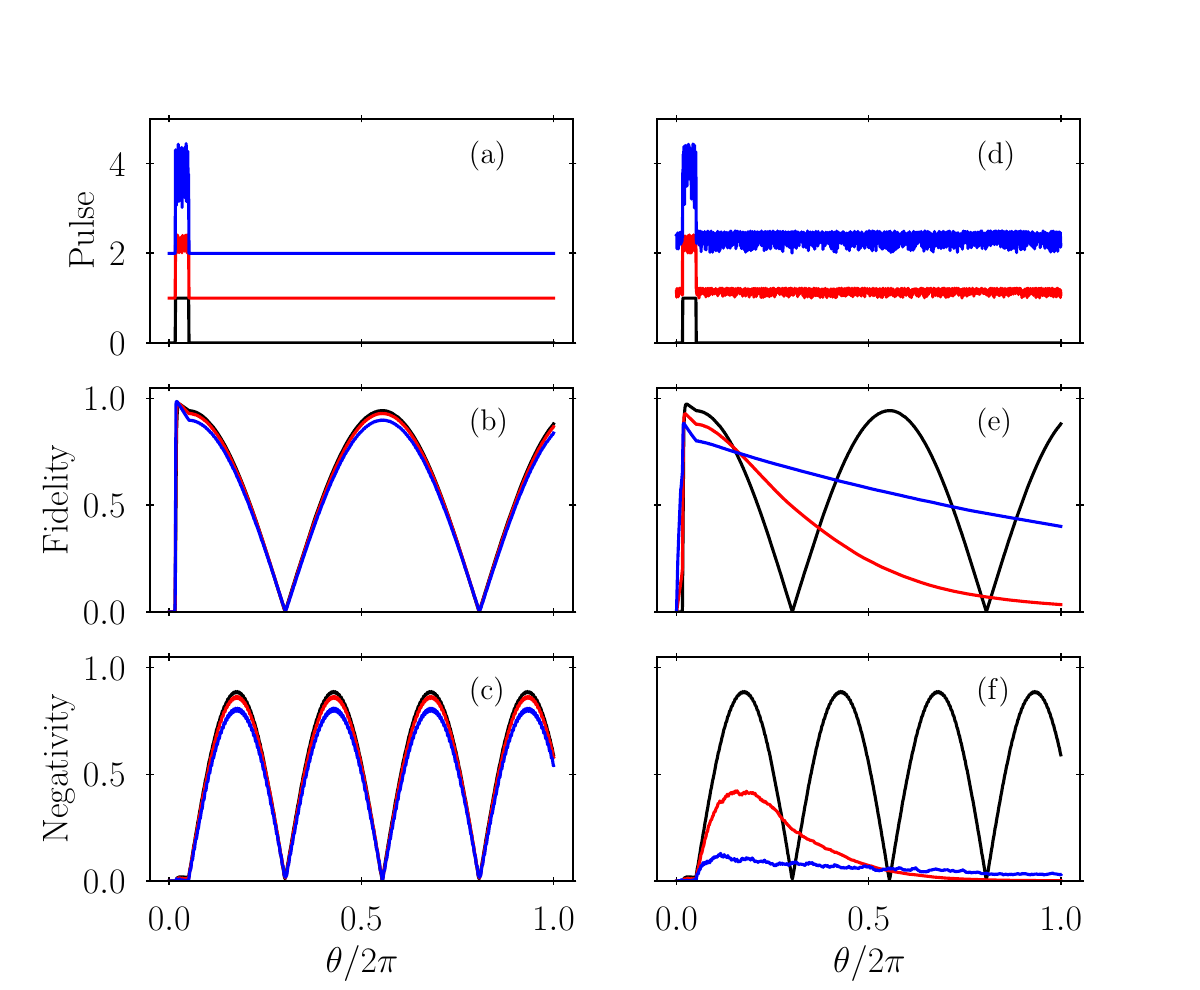}
	\caption{(Color online) Quantum dynamics under amplitude noise applied to a square pulse driving the system. Panels (a)–(c) depict the case where noise is restricted to the pulse width. 
(a) Noisy pulse profiles with three amplitudes: black (low), red (moderate, shown with an upward offset of 1), and blue (high, shown with an upward offset of 2). (b) Fidelity and (c) negativity as functions of time reveal increasing suppression with higher noise amplitudes. Once the pulse ends, the amplitudes of fidelity and negativity remain stable, reflecting the noise-free nature of the subsequent evolution.
Panels (d)–(f) correspond to scenarios where noise persists throughout the entire quantum evolution. (d) Pulse profiles under continuous noise, (e) fidelity, and (f) negativity show pronounced degradation over time. In the blue curve (high noise amplitude), negativity fails to exceed 0.1, indicating that entanglement is not established at any point in the dynamics.\label{fig_pulse_with_noise}}
\end{figure}

To continuing emulate more realistic experimental conditions, amplitude noise is introduced to the square pulse driving the system. This allows us to assess the sensitivity of entanglement formation to external fluctuations and gain deeper insights into protocol performance under noisy environments commonly encountered in practice. To isolate the effect of amplitude noise, the phase flip error discussed in Fig. (\ref{fig5}) is deactivated in this analysis.
Figure (\ref{fig_pulse_with_noise}) presents the quantum dynamics resulting from a square pulse initialization at three distinct noise amplitudes. In panels (a), (b), and (c), noise is confined within the pulse duration. Panel (a) shows the noisy pulse profile; panels (b) and (c) depict the corresponding fidelity and negativity as functions of time. As the fluctuation amplitude increases, fidelity with respect to the target entangled state diminishes, and negativity is significantly suppressed.
In panels (d), (e), and (f), we explore the more detrimental scenario in which electronic noise is present throughout the entire evolution, not just during the pulse. Here, increasing fluctuation amplitude leads to a complete destruction of entanglement.

Our results demonstrate that both phase flip errors and amplitude fluctuations in the driving pulse can severely compromise the formation and preservation of entanglement. While phase flip errors lead to rapid dephasing, electronic noise, whether confined to the pulse or persistent throughout the evolution, significantly degrades fidelity and suppresses quantum correlations. Together, these findings underscore the importance of robust control and error mitigation strategies for maintaining entanglement in realistic environments.

A final comment is necessary, as the predominant focus in implementing quantum computing on semiconductor platforms has been directed toward spin qubits~\cite{loss1998,watson2018,mills2022,ciorga2001,elzerman2004,connors20}. This preference is largely attributed to the relatively long decoherence times of spin qubits—typically on the order of microseconds~\cite{meunier25}—which significantly exceed the entanglement lifetime of the state proposed in this work, which follows Fujisawa and co-workers platform. Nevertheless, the charge qubit examined here offer some advantages: it enables straightforward scalability of entangled states~\cite{nogueira21} and allows faster control via electric fields, facilitating easier integration with other systems~\cite{zhang22}. New fabrication paths are recently reported~\cite{zhou23}, which promised longer coherence timescale.

\section{Concluding remarks}
\label{sec:conc}
We have shown that quantum entanglement in charge qubits can be enhanced by properly tuning the initialization gate pulses. In particular, shorter pulses and stronger lead-dot coupling result in more effective initialization, leading to high-fidelity entangled states. We explored distinct pulse profiles, specifically square and Gaussian shapes, to understand how temporal structure fundamentally affects entanglement formation. Our simulations indicate that states with fidelity exceeding 90\% and negativity above 0.8 can be achieved under optimized conditions. The linear entropy increases during initialization, remaining finite when the initialization pulse is turned off, with higher values for broader pulses. Several mechanisms that induce dephasing and noise, such as phase-flip errors and amplitude fluctuations in the gate voltage, were introduced to emulate realistic experimental conditions. Our results demonstrate that these noise sources significantly suppress entanglement, especially when electronic fluctuations persist beyond the pulse duration. This highlights the sensitivity of quantum coherence to environmental disturbances and underscores the need for robust, noise-resilient control strategies. Our findings provide insights into controlling quantum correlations in mesoscopic systems and emphasize the critical importance of error mitigation in the development of high-performance electronic quantum devices.

\begin{acknowledgments}
E. M. Fernandes acknowledges CAPES and FAPEMIG for financial support. F. M. Souza and L. Sanz thank CNPq for financial support (No. 422350/2021-4). L. Sanz also thanks the Brazilian National Institute of Science and Technology of Quantum Information (grant CNPq No. 465469/2014-0).
\end{acknowledgments}


\begin{thebibliography}{10}
\bibitem{nielsenchuang2007} M. A. Nielsen and I. L. Chuang,  Quantum Computation and Quantum Information, 1st. Ed., Foundation Books (2007).


\bibitem{shor1994} P. W. Shor, Algorithms for quantum computation: discrete logarithms and factoring, Proc. 35th Annual Symposium on Foundations of Computer Science, 124 (IEEE, 1994).

\bibitem{shor1999} P. W. Shor, Polynomial-time algorithms for prime factorization and discrete logarithms on a quantum computer. SIAM Rev. \textbf{41}, 303 (1999).

\bibitem{grover1996} L. K. Grover, A fast quantum mechanical algorithm for database search. Proc. of the Twenty-Eighth Annual ACM Symposium on Theory of Computing. New York: Association for Computing Machinery, 212 (1996).

\bibitem{wiebe2012} N. Wiebe, D. Braun, and S. Lloyd, Quantum algorithm for data fitting, Phys. Rev. Lett. \textbf{109}, 050505 (2012).
\bibitem{biamonte2017} J. Biamonte, P. Wittek, N. Pancotti, P. Rebentrost, N. Wiebe, and S. Lloyd, Quantum machine learning, Nature \textbf{549}, 195 (2017).
\bibitem{leon2021} N. P. de Leon, K. M. Itoh, D. Kim, K. K. Mehta, T. E. Northup, H. Paik, B. S. Palmer, N. Samarth, S. Sangtawesin, and D. W. Steuerman, Materials challenges and opportunities for quantum computing hardware, Science \textbf{372}, eabb2823 (2021).
\bibitem{arute2019}  F. Arute, K. Arya, R. Babbush, D. Bacon, J. C. Bardin, R. Barends, R. Biswas, S. Boixo, F. G. S. L. Brandao, D. A. Buell, B. Burkett, Y. Chen, Z. Chen, B. Chiaro, R. Collins, W. Courtney, A. Dunsworth, E. Farhi, B. Foxen, A. Fowler, C. Gidney, M. Giustina, R. Graff, K. Guerin, S. Habegger, M. P. Harrigan, M. J. Hartmann, A. Ho, M. Hoffmann, T. Huang, T. S. Humble, S. V. Isakov, E. Jeffrey, Z. Jiang, D. Kafri, K. Kechedzhi, J. Kelly, P. V. Klimov, S. Knysh, A. Korotkov, F. Kostritsa, D. Landhuis, M. Lindmark, E. Lucero, D. Lyakh, S. Mandrà, J. R. McClean, M. McEwen, A. Megrant, X. Mi, K. Michielsen, M. Mohseni, J. Mutus, O. Naaman, M. Neeley, C. Neill, M. Y. Niu, E. Ostby, A. Petukhov, J. C. Platt, C. Quintana, E. G. Rieffel, P. Roushan, N. C. Rubin, D. Sank, K. J. Satzinger, V. Smelyanskiy, K. J. Sung, M. D. Trevithick, A. Vainsencher, B. Villalonga, T. White, Z. J. Yao, P. Yeh, A. Zalcman, H. Neven, and J. M. Martinis, Quantum supremacy using a programmable superconducting processor, Nature \textbf{574}, 505 (2019). 

\bibitem{ykim2023} Y. Kim, A. Eddins, S. Anand, K. X. Wei, E. van den Berg, S. Rosenblatt, H. Nayfeh, Y. Wu, M. Zaletel, K. Temme, and A. Kandala, Evidence for the utility of quantum computing before fault tolerance, Nature \textbf{618}, 500 (2023).

\bibitem{vandamme2024} J. Van Damme, S. Massar, R. Acharya, Ts. Ivanov, D. Perez Lozano, Y. Canvel, M. Demarets, D. Vangoidsenhoven, Y. Hermans, J. G. Lai, A. M. Vadiraj, M. Mongillo, D. Wan, J. De Boeck, A. Poto\v{c}nik and K. De Greve, Advanced CMOS manufacturing of superconducting qubits on 300 mm wafers, Nature \textbf{634}, 74 (2024).




\bibitem{Rower2024} D. A. Rower, L. Ding, H. Zhang, M. Hays, J. An, P. M. Harrington, I. T. Rosen, J. M. Gertler, T. M. Hazard, B. M. Niedzielski, M. E. Schwartz, S. Gustavsson, K. Serniak, J. A. Grover, and W. D. Oliver, Suppressing counter-rotating errors for fast single-qubit gates with fluxonium, PRX Quantum \textbf{5}, 040342 (2024).

\bibitem{zwerver2022} A. M. J. Zwerver, T. Kr\"ahenmann, T. F. Watson, L. Lampert, H. C. George, R. Pillarisetty, S. A. Bojarski, P. Amin, S. V. Amitonov, J. M. Boter, R. Caudillo, D. Correas-Serrano, J. P. Dehollain, G. Droulers, E. M. Henry, R. Kotlyar, M. Lodari, F. Lüthi, D. J. Michalak, B. K. Mueller, S. Neyens, J. Roberts, N. Samkharadze, G. Zheng, O. K. Zietz, G. Scappucci, M. Veldhorst, L. M. K. Vandersypen, and J. S. Clarke, Qubits made by advanced semiconductor manufacturing, Nature Electronics \textbf{5}, 184 (2022).

\bibitem{chatterjee2021} A. Chatterjee, P. Stevenson, S. De Franceschi, A. Morello, N. P. de Leon, and F. Kuemmeth,  . Semiconductor qubits in practice. Nature Reviews Physics \textbf{3}, 157 (2021)

\bibitem{Zhang2019} X. Zhang, H. O. Li, G. Cao, M. Xiao, G. C. Guo, and G. P. Guo, Semiconductor quantum computation, National Science Review \textbf{6}, 32 (2019).

\bibitem{levy2002} J. Levy, Universal quantum computation with spin-1/2 pairs and Heisenberg exhange, Phys. Rev. Lett. \textbf{89}, 147902 (2002).

\bibitem{divincenzo2000} D. P. DiVincenzo, D. Bacon, J. Kempe, G. Burkard, and K. B. Whaley, Universal computation with the exchange interaction, Nature \textbf{408}, 339 (2000).

\bibitem{hayashi2003} T. Hayashi1, T. Fujisawa, H. D. Cheong, Y. H. Jeong, and Y. Hirayama, Coherent Manipulation of Electronic States in a Double Quantum Dot, Phys. Rev. Lett. \textbf{91}, 226804 (2003).


\bibitem{shinkai2007} G. Shinkai, T. Hayashi, Y. Hirayama, and T. Fujisawa, Controlled resonant tunneling in a coupled double-quantum-dot system, Appl. Phys. Lett. \textbf{90}, 103116 (2007).

\bibitem{shinkai2009a} G. Shinkai, T. Hayashi, T. Ota, and T. Fujisawa, Correlated coherent oscillations in coupled semiconductor charge qubits, Phys. Rev. Lett. \textbf{103}, 056802 (2009).

\bibitem{peterson2010} K. D. Petersson, J. R. Petta, H. Lu, and A. C. Gossard, Quantum Coherence in a One-Electron Semiconductor Charge Qubit, Phys. Rev. Lett. \textbf{105}, 246804 (2010)

\bibitem{dkim2015}   D. Kim, D. R. Ward, C. B. Simmons, J. K. Gamble, R. Blume-Kohout, E. Nielsen, D. E. Savage, M. G. Lagally, M. Friesen, S. N. Coppersmith, and M. A. Eriksson, Microwave-driven coherent operation of a semiconductor quantum dot charge qubit, Nat. Nanotechnol. \textbf{10}, 243 (2015).


\bibitem{takeda2016} K. Takeda, J. Kamioka, T. Otsuka, J. Yoneda, T. Nakajima, M. R. Delbecq, S. Amaha, G. Allison, T. Kodera, S. Oda, and S. Tarucha, A fault-tolerant addressable spin qubit in a natural silicon quantum dot, Sci. Adv. \textbf{2}, e1600694 (2016).


\bibitem{yoneda2018} J. Yoneda, K. Takeda, T. Otsuka, T. Nakajima, M. R. Delbecq, G. Allison, T. Honda, T. Kodera, S. Oda, Y. Hoshi, N. Usami, K. M. Itoh, and S. Tarucha, A quantum-dot spin qubit with coherence limited by charge noise and fidelity higher than 99.9\%, Nat. Nanotechnol. \textbf{13}, 102 (2018).

\bibitem{zajac2018}D. M. Zajac, A. J. Sigillito, M. Russ, F. Borjans, J. M. Taylor, G. Burkard, and J. R. Petta, Resonantly driven CNOT gate for electron spins, Science \textbf{359}, 439 (2018).

\bibitem{BURKARD2023}  G. Burkard, T. D. Ladd, A. Pan, J. M. Nichol, and J. R. Petta, Semiconductor spin qubits, Rev. Mod. Phys. \textbf{95}, 025003 (2023).


\bibitem{loss1998} D. Loss and D. P. DiVincenzo, Quantum computation with quantum dots, Phys. Rev. A \textbf{57}, 120 (1998).

\bibitem{watson2018} T. F. Watson, S. G. J. Philips, E. Kawakami, D. R. Ward, P. Scarlino, M. Veldhorst, D. E. Savage, M. G. Lagally, M. Friesen, S. N. Coppersmith, M. A. Eriksson, and L. M. K. Vandersypen, A programmable two-qubit quantum processor in silicon, Nature \textbf{555}, 633 (2018).


\bibitem{mills2022}  A. R. Mills, C. R. Guinn, M. J. Gullans, A. J. Sigillito, M. M. Feldman, E. Nielsen, and J. R. Petta, Two-qubit silicon quantum processor with operation fidelity exceeding 99\%, Sci. Adv. \textbf{8}, eabn5130 (2022).


\bibitem{ciorga2001} M. Ciorga, A. S. Sachrajda, P. Hawrylak, C. Gould, P. Zawadzki, Y. Feng, and Z. Wasilewski, Readout of a single electron spin-based quantum bit by current detection, Phys. E \textbf{11}, 35 (2011)


\bibitem{fujisawa2002} T. Fujisawa, D. G. Austing, Y. Tokura, Y. Hirayama, and S. Tarucha, Allowed and forbidden transitions in artificial hydrogen and helium atoms, Nature \textbf{419}, 278 (2002).


\bibitem{elzerman2004} J. M. Elzerman, R. Hanson, L. H. W. van Beveren, B. Witkamp, L. M. K. Vandersypen, and L. P. Kouwenhoven, Single-shot read-out of an individual electron spin in a quantum dot, Nature \textbf{430}, 431 (2004).

\bibitem{connors20} E. J. Connors, J. Nelson, L. F. Edge, and J. M. Nichol, Charge-noise spectroscopy of Si/SiGe quantum dots via dynamically-decoupled exchange oscillations, Nat. Comm. \textbf{13}, 940 (2022).

\bibitem{macquarrie20} E. R. MacQuarrie, S. F. Neyens, J. P. Dodson, J. Corrigan, B. Thorgrimsson, N. Holman, M. Palma, L. F. Edge, M. Friesen, S. N. Coppersmith, and M. A. Eriksson, Progress toward a capacitively mediated {CNOT} between two charge qubits in {Si/SiGe}, Quantum Information \textbf{6}, 81 (2020).

\bibitem{shi2013} Z. Shi, C. B. Simmons, D. R. Ward, J. R. Prance, R. T. Mohr, T. S. Koh, J. K. Gamble, X. Wu, D. E. Savage, M. G. Lagally, M. Friesen, S. N. Coppersmith, and M. A. Eriksson, Coherent quantum oscillations and echo measurements of a Si charge qubit, Phys. Rev. B \textbf{88}, 075416 (2013).

\bibitem{UDDIN2022} W. Uddin, B. Khan, S. Dewan, and S. Das, Silicon-based qubit technology: progress and future prospects, Bull. Mater. Sci. \textbf{45}, 46 (2022).

\bibitem{divincenzo1995} D. P. DiVincenzo, Two-bit gates are universal for quantum computation, Phys. Rev. A \textbf{51}, 1015 (1995).

\bibitem{fujisawa2011} T. Fujisawa, G. Shinkai, T. Hayashi, and T. Ota,  Multiple two-qubit operations for a coupled semiconductor charge qubit. Physica E: Low-dimensional Systems and Nanostructures \textbf{43}, 730 (2011).

\bibitem{oliveira2015} P. A. Oliveira and L. Sanz, Bell states and entanglement dynamics on two coupled quantum molecules, Ann. Phys. \textbf{356}, 244 (2015).

\bibitem{souza2017}  F. M. Souza and L. Sanz, Lindblad formalism based on fermion-to-qubit mapping for nonequilibrium open quantum systems, Phys. Rev. A \textbf{96}, 052110 (2017).

\bibitem{valente10} D. C. B. Valente, E. R. Mucciolo, and F. K. Wilhelm, Decoherence by electromagnetic fluctuations in double-quantum-dot charge qubits, Phys. Rev. B \textbf{82}, 125302 (2010).

\bibitem{vorojtsov05} S. Vorojtsov, E. R. Mucciolo, and H. U. Baranger, Phonon decoherence of a double quantum dot charge qubit, Phys. Rev. B \textbf{71}, 205322 (2005).

\bibitem{wingreen1993} N. S. Wingreen, A. P. Jauho, and Y. Meir, Time-dependent transport through a mesoscopic structure, Phys. Rev. B \textbf{48}, 8487 (1993).

\bibitem{jauho1994} A. P. Jauho, N. S. Wingreen, and Y. Meir, Time-dependent transport in interacting and noninteracting resonant-tunneling systems, Phys. Rev. B \textbf{50}, 5528 (1994).

\bibitem{platero1} D. {Fernández-Fernández}, Yue Ban, and G. Platero. Quantum control of hole spin qubits in double quantum dots, Phys. Rev. Applied \textbf{18}, 054090 (2022). 

\bibitem{platero2} D. {Fernández-Fernández}, Yue Ban, and G. Platero. Flying spin qubits in quantum dot arrays driven by spin-orbit interaction, Quantum \textbf{8}, 1533 (2024). 

\bibitem{wehbi2020} S. Wehbi, F. El Bassri, D. Pagnoux, P. Leproux, D. Arnaud-Cormos, P. Leveque, A. Bertrand, and V. Couderc,
Generation of kilovolt, picosecond electric pulses by coherent combining in optoelectronic system," Proc. SPIE 11279, Terahertz, RF, Millimeter, and Submillimeter-Wave Technology and Applications XIII, 112791V (2020)

\bibitem{wehbi2022}  S. Wehbi, N. Tabcheh, A. Tonello, R. Orlacchio, P. Lévêque, D. Arnaud-Cormos, T. Mansuryan, M. Fabert, O. Tantot, S. Vergnole, and V. Couderc, Temporal shaping of high-voltage picosecond electric pulses for electronic spectroscopy and bioelectric applications, Microw. Opt. Technol. Lett.   \textbf{65}, 717 (2023).

\bibitem{zhau2014} W. Zhu, D. Xiao, Y. Liu, S. J. Gong, and C. G. Duan, Picosecond electric field pulse induced coherent magnetic switching in MgO/FePt/Pt (001)-based tunnel junctions: a multiscale study. Sci. Rep. \textbf{4}, 4117 (2014).

\bibitem{Horodecki1996a} R. Horodecki and  M. Horodecki. Information-theoretic aspects of inseparability of mixed states. Phys. Rev. A \textbf{54}, 1838 (1996).

\bibitem{Horodecki1996b} M. Horodecki, P Horodecki and R. Horodecki, Separability of mixed states: necessary and sufficient conditions. Phys. Lett.
A \textbf{223} (1996) 1. 

\bibitem{peres1996}  A. Peres, Separability criterion for density matrices. Physical Review Letters \textbf{77}, 1413 (1996).

\bibitem{Horodecki2009} R. Horodecki, P. Horodecki, M. Horodecki, and K. Horodecki, Quantum entanglement, Rev. Mod. Phys. \textbf{81}, 865 (2009).

\bibitem{meunier25} T. Meunier, N. Daval, F. Perruchot, M. Vinet, Silicon spin qubits: a viable path towards industrial
manufacturing of large-scale quantum processors, Eur. Phys. J. A \textbf{61}, 58 (2025).

\bibitem{nogueira21} J. Nogueira, P. A. Oliveira, F. M. Souza, and L. Sanz, Dynamic generation of Greenberger-Horne-Zeilinger states with coupled charge qubits, Phys. Rev. A \textbf{103}, 032438 (2021).

\bibitem{zhang22} C. Zhang, G. Xuan Chan, X. Wang, and Z.-Y. Xue, Coupling two charge qubits via a superconducting resonator operating in the resonant and dispersive regimes, Phys. Rev. A \textbf{106}, 032608 (2022).

\bibitem{zhou23} X. Zhou, X. Li, Q. Chen, G. Koolstra, G. Yang, B. Dizdar, Y. Huang, C. S. Wang, X. Han, X. Zhang, D. I. Shuster, and D. Jin, Electron charge qubit with 0.1 millisecond coherence time, Nat. Phys. \textbf{20}, 116 (2024).


\end{thebibliography}
\end{document}